\documentclass[twocolumn,prl,superscriptaddress,showpacs,amsmath,amssymb]{revtex4}
\usepackage{amsfonts}
\usepackage{bbm}

\usepackage{graphicx}
\usepackage{dcolumn}
\usepackage{bm}

\begin{document}

\title{Topological states and quantized current in helical molecules}
\author{Ai-Min Guo}
\email[]{amguo@hit.edu.cn}
\affiliation{Department of Physics, Harbin Institute of Technology,
Harbin 150001, China}
\author{Qing-Feng Sun}
\email[]{sunqf@pku.edu.cn}
\affiliation{International Center for Quantum Materials, School of Physics, Peking University, Beijing 100871, China}
\affiliation{Collaborative Innovation Center of Quantum Matter, Beijing 100871, China}


\begin{abstract}
We report a theoretical study of electron transport along helical molecules under an external electric field, which is perpendicular to the helix axis of the molecule. Our results reveal that the topological states could appear in single-helical molecule and double-stranded DNA in the presence of the perpendicular electric field. And these topological states guarantee adiabatic charge pumping across the helical molecules by rotating the electric field in the transverse plane and the pumped current at zero bias voltage is quantized. In addition, the quantized current constitutes multiple plateaus by scanning the Fermi energy as well as the bias voltage, and hold for various model parameters, since they are topologically protected against perturbations. These results could motivate further experimental and theoretical studies in the electron transport through helical molecules, and pave the way to detect topological states and quantized current in the biological systems.
\end{abstract}


\maketitle

{\it Introduction.}---Helix structures are ubiquitous both in the biological world\cite{snc,erg,gjc} and for synthetic materials\cite{nt,gpx,ye1}. The electron transport along helical molecules, such as DNA and $\alpha$-helical protein, has been receiving much attention among the scientific communities\cite{ch,gx,sjd,bya,rn,xl,jy,ptr,aan,sss,mpc}, because this subject can enrich our knowledge regarding the electronic properties of low-dimensional systems due to the unique helix structure and provides valuable information for understanding the biological processes in living organisms\cite{gsm,me,ye}. It was reported by direct charge transport experiments that double-stranded DNA (dsDNA) could exhibit fascinating physical phenomena, such as the proximity-induced superconductivity\cite{kay}, the negative differential resistance\cite{jpc}, and the piezoelectric effect\cite{bc}. Additionally, it was shown that both dsDNA and $\alpha$-helical protein can behave as electric field-effect transistors\cite{ykh,mg,mav,rs} and as efficient spin filters\cite{ys,gb,gam1,gr,me1,md,gam2,me2}.

On the other hand, topological insulators are very interesting materials that have an insulating gap in the bulk but possess exotic conducting states on their edge or surface\cite{hmz,qxl}. These conducting edge (surface) states are protected by specified symmetries and are robust against disorders. Since then, the topological quantum states have been attracting intense and growing interest in condensed matter and materials physics. Except for two-dimensional (2D) and three-dimensional materials, the topological states were also observed in a variety of one-dimensional (1D) systems, including photonic quasicrystals\cite{kye}, quasiperiodic optical lattices\cite{llj}, atomic Bose-Einstein condensates\cite{mm,sbk}, and a double Peierls chain\cite{cs}. However, the topological states have not yet been reported in the biological molecules.

In this paper, we find that there exist topological states in both single-helical molecule and the dsDNA molecule under an external electric field, which is perpendicular to the helix axis ($z$-axis). In particular, an adiabatic charge pumping occurs in the two-terminal helical molecular systems by rotating the electric field in the transverse plane ($x$-$y$ plane). When the Fermi energy lies in the bulk gap, the pumped current at zero bias voltage is quantized at $Cef$, with integer $C$ being the Chern number of the system, $e$ the elementary charge, and $f$ the rotational frequency of the electric field. And the direction of the pumped current is reversed if the chirality of the molecule is transformed from the right-handed species to the left-handed one. Besides, the quantized current forms several plateaus by sweeping the Fermi energy or the bias voltage, and hold for a very wide range of model parameters. These results suggest that the helical molecules provide a natural platform to explore topological states and quantized current in the 1D systems.

{\it Model.}---The electron transport through the two-terminal helical molecule can be simulated by the Hamiltonian: ${\cal H}={\cal H}_m + {\cal H}_{el}.$ Here, ${\cal H}_m$ is the molecular Hamiltonian as discussed later. ${\cal H}_{el}=\sum_{\beta,k}[\varepsilon_{\beta k}a_{\beta k}^\dagger a_{\beta k} + \sum_{j=1}^Jt_\beta (a_{\beta k}^\dagger c_{jn_{\beta}}+ {\rm H.c.} )]$ describe the left and right metal electrodes and their couplings to the molecule, with $\beta={\rm L}/{\rm R}$ and $n_{{\rm L}/{\rm R}}=1/N$. $a_{\beta k}/c_{jn_{\beta}}$ is the annihilation operator in the electrodes/molecule, $J$ is the number of helical chains, and $N$ is the molecular length.

{\it Single-helical molecule.}---We first consider the simple case of $J=1$, i.e., the single-helical molecule. Then, the molecular Hamiltonian is\cite{gam2,bya1,sct}
\begin{equation}
{\cal H}_m^{(J=1)}= \sum_{n=1}^N \varepsilon_n c_n^{\dagger}c_n+ \sum_{n=1}^{N-1} t c_n^{\dagger} c_{n+1} +{\rm H.c.},\label{eq1}
\end{equation}
where $c_n$ annihilates an electron at site $n$ with the potential energy $\varepsilon_n$ and the hopping integral $t$ is set to the energy unit. In the presence of an external electric field $E_{\rm g}$ which is perpendicular to the helix axis ($z$-axis), the potential energy takes the form\cite{rs,ptr1}
\begin{equation}
\varepsilon_n=eV_{\rm g}  \cos(n\Delta \phi -\varphi).\label{eq2}
\end{equation}
Here, $2V_{\rm g}=2E_{\rm g}R$ is the gate voltage drop across the helical molecule, $R$ is the molecular radius, $\Delta \phi$ is the twist angle between neighboring sites, and the phase $\varphi$ denotes the orientation of the electric field\cite{gam3,ptr1}.

\begin{figure}
\includegraphics[width=8.3cm]{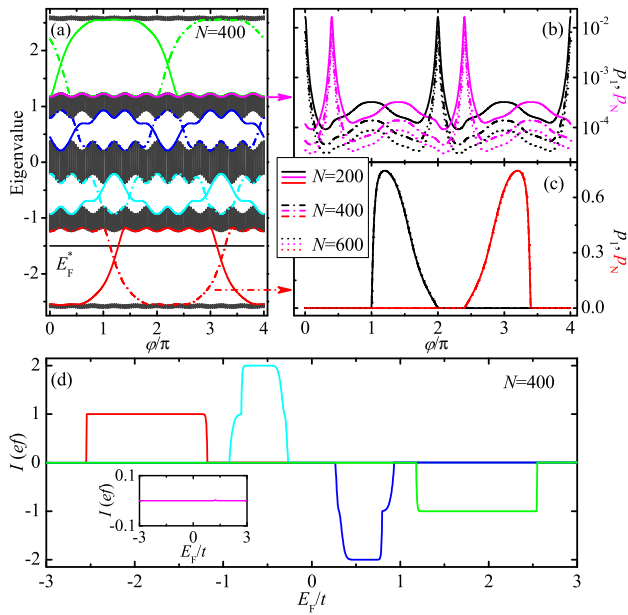}
\caption{\label{fig1} (color online). (a) Energy spectrum of the isolated single-helical molecule under open boundary conditions and the gate voltage. The colored lines across the bulk energy gaps are the topological edge states. Phase-dependent electron densities $p_1$ and $p_N$ at the end sites of the single-helical molecule for (b) a typical bulk state and for (c) a topological edge state with various molecular lengths $N$, as depicted by the arrows. (d) Pumped current $I$ versus Fermi energy $E_F$ for all the topological edge states, where each line corresponds to the edge states with the same color in Fig.~\ref{fig1}(a). The inset shows $I$ for the bulk state which is indicated by the magenta line in Fig.~\ref{fig1}(a). The parameters are $\alpha=1/5$ and $V_{\rm g}=1.5$.}
\end{figure}

The molecular Hamiltonian, Eqs.~(\ref{eq1}) and (\ref{eq2}), is analogous to the Harper-Hofstadter model\cite{hpg,hdr}, which is projected from a 2D square lattice in the presence of a uniform magnetic field, with $\alpha=\Delta\phi/2\pi$ mapping to the magnetic flux threading each primitive cell. Subsequently, it can be seen from the energy spectrum of the single-helical molecule in Fig.~\ref{fig1}(a) that a doublet of edge states traverse each bulk energy gap and intersect at specific values of $\varphi$. These edge states are topologically protected against perturbations until the closure of the bulk gaps and can be characterized by the Chern number\cite{kye,llj}, just as the Hall conductance in a 2D quantum Hall system\cite{tdj}. Within each bulk gap, the topological edge states are localized at the boundaries of the molecule, and thus the electron densities $p_1$ and $p_N$ at the end sites are quite large, as illustrated in Fig.~\ref{fig1}(c). Here, $p_n=|\psi_n|^2$ and $\psi_n$ is the amplitude of the wave function of a specific state at site $n$\cite{sm}. Both $p_1$ and $p_N$ are independent of the molecular length $N$, implying that they are indeed localized at the boundaries and are robust against perturbations. In contrast, the bulk states always lie in the subband region, and $p_1$ and $p_N$ are quite small and decrease with increasing $N$ as $p_n\propto N^{-1}$ (Fig.~\ref{fig1}(b)), due to spatially extended wave function.

The topological edge states can dramatically affect the charge transport along the helical molecule, and a two-terminal setup is considered. Fig.~\ref{fig1}(d) shows pumped current $I$, the current at zero bias voltage, for the single-helical molecule during one pumping cycle, in which the phase $\varphi =2\pi f \tau$ is varied from $0$ to $2\pi$ by rotating the electric field\cite{sm}. Here, $f$ is the rotational frequency with $f=10^6$ Hz and $\tau$ is the time. It clearly appears that the pumped current is quantized at integer multiples of $ef$ and constitutes plateaus by scanning $E_F$, implying that the singe-helical molecule transfers an integer number of electrons (holes) for $E_F<0$ ($E_F>0$). The number of transmitted charges in each pumping cycle equals to that of intersection points between the topological edge states within each bulk gap, i.e., the Chern number of the system. Let us consider the edge state, the red dashed-dot line in Fig.~\ref{fig1}(a), as an example and elucidate the underlying physics of this adiabatic charge pumping. The Fermi energy $E_F ^\star$, the black solid line in Fig.~\ref{fig1}(a), lies in the bottom bulk gap. At $\varphi=0$ ($\tau=0$), the edge state locates above $E_F^\star$ and thus is empty. For a critical $\varphi$ ($\tau$) where the state passes through the Fermi level, an electron will be injected into the edge state from the left electrode and is localized in the left boundary of the molecule. Then, the edge state is occupied. By further increasing $\varphi$ ($\tau$), the electron in the state transports from the left boundary to the right one (Fig.~\ref{fig1}(c)). When the energy of the state surpasses $E_F ^\star$, the electron is transmitted from the edge state to the right electrode and the state becomes empty again. As a result, an electron is pumped across the molecule during two pumping cycles and similar phenomena can be observed regarding the other topological edge states. It is now clear that this adiabatic charge pumping is driven by the topological nature of the edge states and is thus topologically protected against perturbations unless the energy gap closes. While for the bulk state, the pumped current is negligible with the order of magnitude being $10^{-3}$ (the inset of Fig.~\ref{fig1}(d)). In what follows, we focus on the topological edge states, since the bulk states are fragile upon perturbations and their energy range differs from the topological edge states.

\begin{figure}
\includegraphics[width=8.3cm]{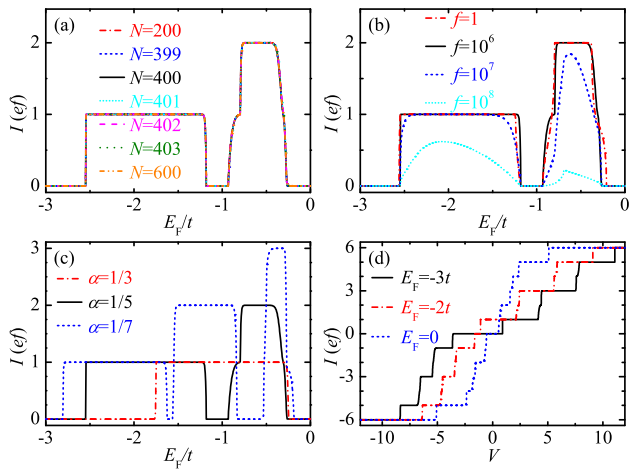}
\caption{\label{fig2} (color online). Pumped current $I$ of the single-helical molecule with (a) different molecular lengths $N$, (b) various rotational frequencies $f$, and (c) several twist angles $\Delta\phi =2\pi \alpha$, where the black solid lines are shown for reference with $\alpha=1/5$, $V_{\rm g}=1.5$, $N=400$, and $f=10^6$ Hz. (d) Current $I$ versus bias voltage $V$ for three $E_F$.}
\end{figure}

To verify the robustness of the topological charge pumping, Figs.~\ref{fig2}(a)-\ref{fig2}(c) plot the pumped current $I$ under a variety of situations. One notices that the $I$-$E_F$ curves are identical to each other for different molecular lengths (Fig.~\ref{fig2}(a)). And the plateaus of pumped current, especially for $I=\pm ef$, can exist in a very wide region of the rotational frequency, ranging from $f=1$ to $10^7$ Hz (Fig.~\ref{fig2}(b)). However, by further increasing $f$, the pumped current will be declined over the entire energy spectrum and hence the plateaus break down, because the speed of electron hopping between the molecule and the electrodes cannot keep up with the rotation rate of the electric field. Besides, the topological charge pumping still holds for different single-helical molecules with various twist angles. The number of plateaus increases with decreasing $\alpha$ and the width becomes narrower when the pumped current is larger (Fig.~\ref{fig2}(c)), since the number and the width of the bulk energy gaps are altered. When the bias voltage $V$ is applied between the left and right electrodes, the current is also quantized at integer multiples of $ef$ and forms several plateaus in the $I$-$V$ curve (Fig.~\ref{fig2}(d)), because of the topological charge pumping. The $I$-$V$ curve is centrally symmetric for $E_F=0$ due to the central symmetry of the $I$-$E_F$ curve (Fig.~\ref{fig1}(d)) and is asymmetric for $E_F\neq0$. Therefore, we conclude that the topological charge pumping is robust, and the plateaus of quantized current can be observed in the $I$-$E_F$ curve at zero bias voltage and in the $I$-$V$ curve at finite bias voltage within a wide range of parameters.

\begin{figure}
\includegraphics[width=8.3cm]{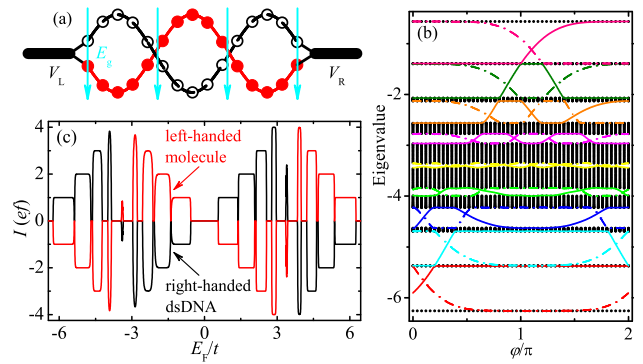}
\caption{\label{fig3} (color online). (a) Sketch of a two-terminal dsDNA under a perpendicular electric field $E_{\rm g}$, where the full and open circles represent the nucleobases. (b) Energy spectrum of the isolated dsDNA under open boundary conditions and the gate voltage. Because of the particle-hole symmetry, the energy spectrum is symmetric with respect to the line $E=0$ and is not shown above $E=0$. (c) Pumped current $I$ versus $E_F$ for the dsDNA and the left-handed molecule at zero bias voltage. The parameters are $V_{\rm g}=1.5$, $N=200$, $f=10^6$ Hz, $\Delta \varepsilon=3t$, and $\lambda=1.5t$.}
\end{figure}

{\it Double-stranded DNA molecule.}---We then study the topological charge pumping in the dsDNA molecule under the electric field $E_{\rm g}$ (Fig.~\ref{fig3}(a)). The Hamiltonian of the dsDNA molecule is\cite{mav,gam1,sct}
\begin{equation}
\begin{aligned}
{\cal H}_m^{(J=2)} =& \sum_{j=1}^2 \sum_{n=1}^N \varepsilon_{jn} c_{jn}^{\dagger} c_{jn} + \sum_{j=1}^2 \sum_{n=1}^{N-1} t c_{jn}^{\dagger} c_{jn+1} \\& +\sum_{n=1}^N \lambda c_{1n}^{\dagger} c_{2n} +{\rm H.c.},\label{eq3}
\end{aligned}
\end{equation}
which can be regarded by coupling two single-helical molecules. The potential energy is\cite{mav,rs,gam3}
\begin{equation}
\varepsilon_{jn}=-(-1)^j[eV_{\rm g}\cos(n\Delta \phi -\varphi)+\Delta \varepsilon],\label{eq4}
\end{equation}
where $\Delta\phi=2\pi/10$, $2\Delta \varepsilon$ is the potential energy difference between the two chains of a homogeneous dsDNA at $V_{\rm g}=0$, and the other parameters are the same as Eq.~(\ref{eq2}). The model parameters are extracted from first-principles calculations with $\Delta \varepsilon=3t$ and $\lambda=1.5t$\cite{yyj,sk,hlgd}.

One can see from Fig.~\ref{fig3}(b) that the lower energy spectrum ($E<0$) of the dsDNA consists of ten subbands as expected. And a pair of topological edge states cross each energy gap and intersect at specific values of $\varphi$. Identical features can be found in the upper energy spectrum ($E>0$), due to the particle-hole symmetry. Consequently, the topological charge pumping can also be observed in the dsDNA that the pumped current is quantized at integer multiples of $ef$ and forms plateaus by scanning $E_F$, as indicated by the black line in Fig.~\ref{fig3}(c). The plateaus locate at the energy gaps and will not be affected by the bulk states. When the plateau is farther away from the lower (upper) band center, it is wider and is more robust because of larger energy gap (Fig.~\ref{fig3}(b)). Besides, the pumped current presents alternating electron-like behavior ($I>0$) and hole-like behavior ($I<0$) in the energy spectrum. When the chirality of the molecule is changed from the right-handed species to the left-handed one by using the reflection symmetry, the direction of the current is inverted (Fig.~\ref{fig3}(c)), i.e., $I(\Delta\phi)= -I(-\Delta\phi)$.

\begin{figure}
\includegraphics[width=8.3cm]{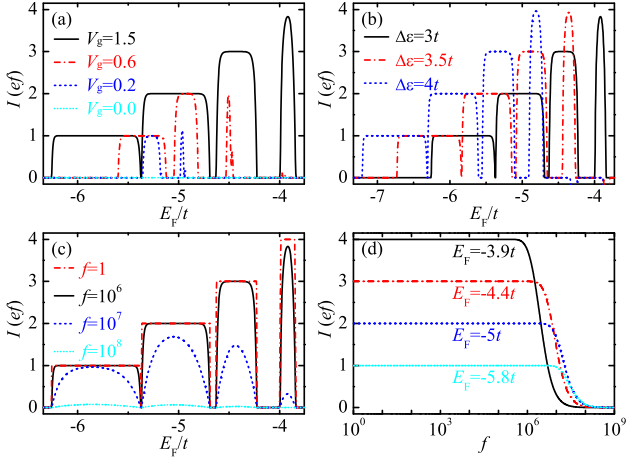}
\caption{\label{fig4} (color online). Pumped current $I$ of the dsDNA with (a) several $V_{\rm g}$, (b) three $\Delta \varepsilon$, and (c) different $f$, where the other parameters are the same as Fig.~\ref{fig3}(c). (d) $I$ versus $f$ for the dsDNA with several $E_F$.}
\end{figure}

Figs.~\ref{fig4}(a)-\ref{fig4}(d) display the pumped current of the dsDNA under various model parameters. For $V_{\rm g}=0$, all the electronic states are trivial and thus the pumped current is zero (the cyan dotted line in Fig.~\ref{fig4}(a)). When the gate voltage is implemented, the topological states emerge and give rise to plateaus of quantized current, even for small gate voltage of $V_{\rm g}=0.2$ (the blue dashed line in Fig.~\ref{fig4}(a)). By increasing $V_{\rm g}$, the plateaus of small $I$ appear first and the original ones broaden. As $\Delta \varepsilon$ rises, the $I$-$E_F$ curve is translated toward lower energy and the conformation of the plateaus keeps almost unchanged (Fig.~\ref{fig4}(b)). Additionally, the plateaus of pumped current can occur in a very wide range of $f$ (Fig.~\ref{fig4}(c)) and that of smaller $I$ are more robust against fast rotation of the electric field (Fig.~\ref{fig4}(d)). Further studies reveal that the plateaus of pumped current remain in the dsDNA by varying, e.g., the interchain hopping integral $\lambda$ and the molecular length. Therefore, we conclude that the topological charge pumping of the dsDNA arises from its intrinsic helix structure, and the quantized pumped current, especially for $I=\pm ef$, is robust and may be detected in the experiments.

\begin{figure}
\includegraphics[width=8.3cm]{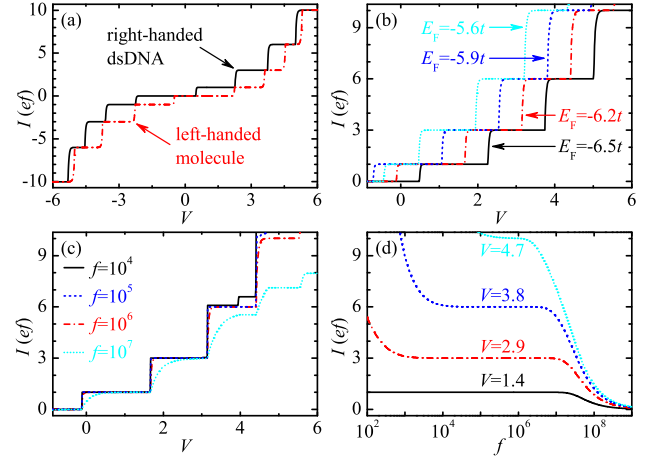}
\caption{\label{fig5} (color online). $I$-$V$ curves for (a) the dsDNA and the left-handed molecule with $E_F=-6.5t$, for the dsDNA with (b) several $E_F$ and (c) various $f$ by fixing $E_F=-6.2t$. (d) $I$ versus $f$ for the dsDNA with $E_F=-6.2t$. The other parameters are the same as Fig.~\ref{fig3}(c).}
\end{figure}

Finally, we study the current of the dsDNA at finite bias voltage, as shown in Figs.~\ref{fig5}(a)-\ref{fig5}(d). Because of the topological charge pumping, the plateaus of quantized current emerge at integer multiples of $ef$ for both the dsDNA and the left-handed molecule, and each $I$-$V$ curve is asymmetric (Fig.~\ref{fig5}(a)). For instance, the current of the dsDNA is $I=ef$ for $V\in (0.6,2.1)$ and vanishes for $V\in (-2.1,-0.6)$, implying that the dsDNA can serve as a molecular switch by reversing the bias voltage. When the Fermi energy is increased, the $I$-$V$ curve is shifted toward lower bias voltage and the plateaus are almost the same (Fig.~\ref{fig5}(b)). It is interesting that the current can also flow across the molecule from the left electrode to the right one even if the chemical potential of the right electrode is higher ($\mu_{\rm R}>\mu_{\rm L}$), i.e., $I$ is positive at $V<0$ (Fig.~\ref{fig2}(d) and Fig.~\ref{fig5}(b)). This arises from the topological charge pumping and the quantized current occurs when $\mu_{\rm L}$ and $\mu_{\rm R}$ locate in the same energy gap. Similarly, the plateaus of the $I$-$V$ curve can exist in a wide range of $f$ (Fig.~\ref{fig5}(c)) and that of smaller $I$ is more robust (Fig.~\ref{fig5}(d)). Therefore, the plateaus of quantized current also appear in the $I$-$V$ curve of the dsDNA at finite bias voltage.

{\it Conclusion.}---In summary, we show that the topological states could emerge in the helical molecules, such as single-helical molecule and double-stranded DNA, in the presence of a perpendicular electric field. A topological charge pumping is predicted by rotating the electric field in the transverse plane and integer number $C$ of electrons is pumped through the helical molecules in each pumping cycle, with $C$ the Chern number of the system. Additionally, the quantized current forms multiple plateaus by scanning the Fermi energy or the bias voltage. These plateaus are topologically protected and are of measurable width, and may thus be realized in the charge transport experiments.

This work was supported by NBRP of China (2012CB921303 and 2015CB921102), NSF-China under Grants No. 11274364,  11504066, and 11574007, and FRFCU under Grant No. AUGA5710013615.

\end{document}